%% file: main.tex
\documentclass{article}
\input{macros}

\begin{document}

\input{titlepage}




\newpage
\doublespacing

Use of machine learning to estimate nuisance functions (e.g. outcomes models, propensity score models) in estimators used in causal inference is increasingly common, as it can mitigate bias due to model misspecification. However, it can be challenging to achieve valid inference (e.g., estimate valid confidence intervals). The efficient influence function (EIF) provides a recipe to go from a statistical estimand relevant to our causal question, to an estimator that can validly incorporate machine learning. Our companion paper, Renson et al. 2025,\cite{renson_pulling_2025} provides a thorough but approachable description of the EIF, along with a guide through the steps to go from a unique statistical estimand to development of one type of EIF-based estimator, the so-called one-step estimator. Another commonly used estimator based on the EIF is the targeted maximum likelihood/minimum loss estimator (TMLE).\cite{van_der_laan_targeted_2011,schuler_targeted_2017} Construction of TMLEs is well-discussed in the statistical literature,\cite{van_der_laan_targeted_2006,van_der_laan_targeted_2011,van_der_laan_targeted_2010} but there remains a gap in translation to a more applied audience. In this letter, which supplements Renson et al., we aim to provide a more accessible illustration of how to construct a TMLE. 

\section*{A reminder: score equations}
We start with a brief review of score equations because they will appear in the construction of the TMLEs in the next section. Consider the parametric linear model,
$$\E(Z|X)=\beta_0 + \beta_1 f(X),$$ where $f(X)$ is some transformation of the covariates $X$, e.g., $f(x)=x^2$. Typically we use maximum likelihood to estimate $\boldsymbol{\beta}=(\beta_0,\beta_1)$ parameters. That is, we find the values of $\boldsymbol{\beta}$ that maximize the likelihood (or log-likelihood). Recall from calculus that at the maximum of a function the slope of the tangent line is zero. Thus, equivalent to maximum likelihood estimation is to find the values of $\boldsymbol{\beta}$ where the slope of the tangent to the log-likelihood (i.e., the derivative of the log-likelihood with respect to each parameter) is zero. We call the functions for the derivative of the log-likelihood with respect to each parameter the \textit{score functions}. The score functions for the model above, one for each $\beta$, are 1) $Z_i - \widehat\E(Z|X_i)$ for $\beta_0$ and 2) $f(X_i)(Z_i - \widehat\E(Z|X_i))$ for $\beta_1$ where $i$ indexes $n$ individuals in the sample and $\widehat\E(Z|X_i)=\widehat\beta_0 + \widehat\beta_1 f(X_i)$. Therefore, using maximum likelihood estimation is equivalent finding the parameter values where the sum of each score function over the sample equals zero. We call the sum of each score function over the sample set to zero the \textit{score equations}. The score equations for the example linear model are,
\begin{align}
    0&=\sum_{i=1}^n (Z_i - \widehat\E(Z|X_i))\notag \\
    0&=\sum_{i=1}^n f(X_i)(Z_i - \widehat\E(Z|X_i)) \label{eq:score}
\end{align}
For a model that includes an offset (i.e., a variable with a fixed coefficient) $b_i$, e.g., $\E(Z|X)= b_i + \beta_0 + \beta_1 f(X_i)$, the score functions are $Z_i - (b_i + \widehat \beta_0 + \widehat\beta_1 f(X_i))$ and $f(X_i)(Z_i - (b_i + \widehat \beta_0 + \widehat\beta_1 f(X_i)))$. In the score equations above individuals are weighted equally ($1/n$), but individuals can be weighted differently, e.g., the estimating equation for $\beta_0$ for the example linear model where each individual is weighted by $wt_i$ is \begin{equation}
0=\sum_{i=1}^n wt_i (Z_i - \widehat\E(Z|X_i))\label{eq:scorewt}.
\end{equation}

Although we show them above for linear models, the form of the score equations is the same for any generalized linear model (GLM) with a canonical link, e.g., logit for a logistic model. That is, the GLM $g(\E(Z|X))=\beta_0 + \beta_1 f(X)$ has the same score equations as the linear model except $\widehat\E(Z|X_i)=g^{-1}(\widehat\beta_0 + \widehat\beta_1 f(X_i))$, where $g^{-1}$ is the inverse of the link, e.g., for logit, $g^{-1}(B)=1/(1+\exp(-B))$. 

\section*{Notation}
We use the same notation and example causal parameter as in Renson et al. Let $Y$ be the outcome, $A$ a binary treatment, and $W$ a set of observed covariates. The observed data are independent and identically distributed copies of $O_i=(W_i,A_i,Y_i)$. Our example causal parameter is the expected outcome under no treatment, $\E(Y^0)$, where $Y^0$ is the potential outcome when $A=0$. The causal parameter is identified as $\E(Y^0)=\E(\E(Y|A=0,W))$ under the 
conditional exchangeability, positivity, and causal consistency assumptions. The \textit{plug-in} estimator for $\psi$ is $\psi(\widehat P)=\frac{1}{n}\sum_{i}\widehat\E(Y|A=0,W_i)$, which corresponds to the g-computation estimator. We cannot generally use machine learning to estimate $\widehat \E(Y|A=0,W)$ and obtain theoretically valid inference on $\psi(\widehat P)$ with the plug-in estimator.

As we described in Renson et al., the EIF is a key piece for constructing machine-learning based estimators that allow valid inference. We denote the EIF of $\psi$ as $\phi(O_i,P)$. In Renson et al. we discussed how to derive the EIF under a nonparametric statistical model. 
For $\psi(P)=\E(\E(Y|A=0,W))$, the EIF is given by
$$\phi(O_i,P) = \frac{I(A=0)}{\Pr(A=0|W)}\left[Y-\E(Y|A=0,W)\right] + \E(Y|A=0,W) - \psi(P).$$
Importantly, estimators that solve the estimating equation $\sum_{i=1}^n \phi(O_i, \widehat P) = 0$ are asymptotically linear under certain conditions (e.g., use of cross-fitting, fast enough consistency of regression estimators, etc.),\cite{kennedy_semiparametric_2016} allowing us to obtain valid standard errors with machine learning. Thus, we have a recipe for constructing machine-learning estimators for $\psi$: estimators that solve the estimating equation of the EIF,
{\small
\begin{equation}
0=\sum_{i=1}^n \left\{ \frac{I(A_i=0)}{\widehat\Pr(A=0|W_i)}\left[Y_i-\widehat\E(Y|A=0,W_i)\right] + \widehat\E(Y|A=0,W_i) - \widehat\psi(P)\right\}.\label{eq:eifee}
\end{equation}}

One method for solving (\ref{eq:eifee}) is to directly solve for $\widehat\psi(P)$. We can rearrange (\ref{eq:eifee}) into a closed-form and obtain the one-step estimator, also called the augmented inverse probability weighted estimator (AIPW), 
{\small
\begin{align*}
    \widehat{\psi}_{os}&=\frac{1}{n}\sum_{i=1}^n \left\{ \frac{I(A_i=0)}{\widehat\Pr(A=0|W_i)}\left[Y_i-\widehat\E(Y|A=0,W_i)\right] + \widehat\E(Y|A=0,W_i) \right\}.
\end{align*}}
As shown in Renson et al., one-step estimators are more generally formed by adding $\frac{1}{n}\sum_{i=1}^n \widehat \phi(O_i, \widehat P)$ to the plug-in estimator.\cite{kennedy_semiparametric_2023,hines_demystifying_2022} In the case of our example estimand, that formation aligns with a rearrangement of the EIF estimating equation, but this is not always the case.\cite{hines_demystifying_2022} 

\section*{TMLE}
The EIF estimating equation (\ref{eq:eifee}) is reprinted here with the summation distributed so that there are two pieces,
\begin{equation*}
   0=
   \underbracket{\sum_{i=1}^n\left\{\frac{I(A_i=0)}{\widehat\Pr(A=0|W_i)}\left[Y_i-\widehat\E(Y|A=0,W_i)\right] \right\}} + 
   \underbracket{\sum_{i=1}^n \left\{\widehat\E(Y|A=0,W_i) -\widehat{\psi}\right\}}.
\end{equation*}
Another approach to solving (\ref{eq:eifee}) is to find a solution where each piece equals zero. This arrangement shows that the EIF estimating equation can be represented as two estimating equations,
\begin{align}
    0&=\sum_{i=1}^n \frac{I(A_i=0)}{\widehat\Pr(A=0|W_i)}\left[Y_i-\widehat\E^*(Y|A=0,W_i)\right]\label{eq:eeupdate} 
    \\
    0&=\sum_{i=1}^n \left\{\widehat\E^*(Y|A=0,W_i) -\widehat{\psi}\right\} \label{eq:eemean}
\end{align}
Here we use $\widehat{\E}^*(Y|A=0,W_i)$ instead of $\widehat{\E}(Y|A=0,W_i)$ to distinguish it from the nuisance parameter in the plug-in and one-step estimators (i.e, predictions from the $W$-conditional outcome model). We call $\widehat{\E}^*(Y|A=0,W_i)$ a targeted version of $\widehat{\E}(Y|A=0,W_i)$. Equation (\ref{eq:eemean}) can be rearranged into a closed-form estimator of our estimand that is the mean of the targeted outcome predictions, $\widehat \psi_{target} = \frac{1}{n}\sum_{i=1}^n\widehat{\E}^*(Y|A=0,W_i)$. Thus our task is to derive an estimator for $\widehat{\E}^*(Y|A=0,W_i)$ that solves (\ref{eq:eeupdate}). 

Notice that (\ref{eq:eeupdate}) is of the same form as (\ref{eq:score}) with $f(X_i)$ equal to $\frac{I(A_i=0)}{\widehat\Pr(A=0|W_i)}$. Thus we can solve (\ref{eq:eeupdate}) by fitting a GLM of $Y$ on $\frac{I(A_i=0)}{\widehat\Pr(A=0|W_i)}$ (the ``clever" covariate) using maximum likelihood. We call this GLM the targeting model since it is an estimator of $\widehat{\E}^*(Y|A=0,W_i)$. There are many possible targeting models that would solve the above score equation.\cite{vansteelandt_analysis_2010,robins_comment_2007,kang_demystifying_2007} What defines TMLEs as a class of estimators is that the targeting model is specified as an update to an initial estimator $\widehat{\E}(Y|A=0,W_i)$. This is advantageous because $\widehat{\E}(Y|A=0,W_i)$ can be estimated using machine-learning even though the targeting model is estimated by maximum likelihood (only maximum likelihood estimation guarantees solving the score equation (\ref{eq:eeupdate})). We can define 
\[\widehat{\E}^*(Y|A=0,W_i)=\widehat{\E}(Y|A=0,W_i) + \widehat \delta \frac{I(A_i=0)}{\widehat\Pr(A=0|W_i)}.\] Plugging this into (\ref{eq:eeupdate}), we get the score equation $0=\sum_{i=1}^n \frac{I(A_i=0)}{\widehat\Pr(A=0|W_i)}\left[Y_i-\left\{\widehat{\E}(Y|A=0,W_i) + \widehat \delta \frac{I(A_i=0)}{\widehat\Pr(A=0|W_i)}\right\}\right]$. Thus, one targeting model is a linear model of $Y_i$ on $\frac{I(A_i=0)}{\widehat\Pr(A=0|W_i)}$ where $\widehat{\E}(Y|A=0,W_i)$ is an offset and there is no intercept. 

Alternatively, notice that (\ref{eq:eeupdate}) is of the same form as (\ref{eq:scorewt}) with $wt_i$ equal to $\frac{I(A_i=0)}{\widehat\Pr(A=0|W_i)}$. We can define 
\[\widehat{\E}^*(Y|A=0,W_i)=\widehat{\E}(Y|A=0,W_i) + \widehat \gamma \] and plug this into (\ref{eq:eeupdate}) to get the score equation $0=\sum_{i=1}^n \frac{I(A_i=0)}{\widehat\Pr(A=0|W_i)}\left[Y_i-\left\{\widehat{\E}(Y|A=0,W_i) + \widehat \gamma\right\}\right]$. Thus, an alternative targeting model is an intercept-only linear model of $Y_i$ where $\widehat{\E}(Y|A=0,W_i)$ is an offset and individuals are weighted by $\frac{I(A_i=0)}{\widehat\Pr(A=0|W_i)}$.  

Here we have presented linear targeting models for simplicity, but using a logistic model is recommended (even for continuous bounded $Y$) because it ensures bounding of $\widehat\E^*(Y|A=0,W_i)$, resulting in better finite sample properties.\cite{gruber_targeted_2010} 
For instance, we can define $\widehat{\E}^*(Y|A=0,W_i)=\text{expit}\left( \text{logit}\left(\widehat{\E}(Y|A=0,W_i)\right) + \widehat\gamma\right)$. Estimating the intercept-only logistic model for $Y_i$ with weights $\frac{I(A_i=0)}{\widehat\Pr(A=0|W_i)}$ can be shown to solve
\begin{equation*}
0=\sum_{i=1}^n \frac{I(A_i=0)}{\widehat{\Pr}(A=0|W_i)}\left[Y_i-\text{expit}\left( \text{logit}\left(\widehat{\E}(Y|A=0,W_i)\right) + \widehat\gamma\right)\right]. 
\end{equation*}

Deriving TMLEs for other estimands may differ in complexity. In the appendix we provide an additional example for the causal parameter of the expected outcome under sustained no treatment (at two time points). In this example the EIF-estimating equation is presented as three estimating equations, two of which are solved by specifying targeting models.

\section*{Discussion} 
Here we illustrated how to construct TMLEs from the estimating equation of the EIF. The one-step estimator is also constructed from the EIF. Although TMLEs and one-step estimators are asymptotically equivalent, they have different finite sample properties. Importantly, TMLEs produce estimates that respect the parameter space of $Y$. This is because the targeting outcome model ensures bounding of $\widehat\E^*(Y|A=0,W_i)$ and the estimate of our estimand is the sample mean of the bounded $\widehat\E^*(Y|A=0,W_i)$. This is not the case with the one-step estimator, which may produce estimates outside the parameter space of $Y$. Bounding produces superior finite sample performance of TMLEs compared to the one-step estimator.\cite{robins_comment_2007} However, for some estimands, TMLEs may be notably more difficult to construct than one-step estimators.

In Renson et al., we discussed rate double robustness, the property that allows an estimator to accommodate machine learning. We can leverage the same work used for the one-step estimator in that paper to show that TMLEs are also rate double robust. Briefly, by setting the mean of the estimated EIF to zero, TMLEs remove part of the first-order error (specifically, the ``drift term''). The analysis of the remaining  error follows steps similar to those presented in Renson et al.: when sample splitting is used, the error converges to a mean zero normal distribution plus a remainder term equal to a product of errors of nuisance functions (thus illustrating rate double robustness). The only difference with the analysis in Renson et al. is that the empirical process term needs careful handling since the targeting model is estimated in the entire sample. Intuitively, EIF-based estimators remove the portion of the error in the plug-in estimator that has the potential to ``blow up" with slow-converging nuisance estimators. One-step estimators and TMLEs remove this portion in distinct ways leading to different properties.

\newpage
\printbibliography

\newpage
\input{appendix}
\end{document}

%% file: macros.tex
\usepackage{amsmath}
\usepackage{graphicx}
\usepackage{amssymb}
\usepackage{fullpage}
\usepackage{enumitem}
\usepackage{setspace}
\usepackage[english]{babel}
\usepackage[titletoc,title]{appendix}
\usepackage{comment}
\usepackage{longtable}
\usepackage{booktabs}
\usepackage{lscape}
\usepackage{xcolor}
\usepackage{placeins}
\usepackage{annotate-equations}
\usepackage{tikz}

\newcommand{\E}{\text{E}}
\newcommand{\I}{\text{I}}

\usepackage{yhmath}
\usepackage{mathtools}
\usepackage{biblatex}

%% file: titlepage.tex

\noindent Article Type: Research Letter\\~\\

\noindent Title: Constructing targeted minimum loss/maximum likelihood estimators: a simple illustration to build intuition\\~\\

\noindent Authors: Rachael K. Ross, Lina M. Montoya, Dana E. Goin, Iv\'an D\'iaz, Audrey Renson\\~\\

\noindent Corresponding author: Dr. Rachael Ross, Department of Epidemiology, Mailman School of Public Health, Columbia University, 722 W 168th St, New York, NY 10032; rachael.k.ross@columbia.edu\\~\\

\noindent Affiliations: Department of Epidemiology, Columbia University, New York, NY (Ross, Goin); Department of Population Health, New York University, New York, NY (Renson); School of Data Science and Society and Department of Biostatistics, University of North Carolina at Chapel Hill, Chapel Hill, NC (Montoya); Department of Population Health, Grossman School of Medicine, New York University, New York, NY (D\'iaz)\\~\\

\noindent Conflicts: None\\~\\

\noindent Funding: AR is supported by a gift from the Bezos foundation. DEG is supported by the National Institutes of Health under award number R00ES033274. LMM is supported by the National Institutes of Health under award number R00MH133985.\\~\\

\noindent Data Availability: NA\\~\\

\noindent Word count: 1441/1500



%% file: appendix.tex

{\LARGE{\center{Appendix for\\~\\
Constructing targeted minimum loss/maximum likelihood estimators: a simple illustration to build intuition}}}

\section*{Additional example}
Treatment is considered at two time points, $A_0$ and $A_1$. $W_0$ and $W_1$ are covariates measured prior to treatment at baseline and at the second time point, respectively. $Y$ is measured after the second time point. There is no loss to follow up. We consider the estimand $\theta = \E(\E(\E(Y|W_0,A_0=0,W_1,A_1=0)|W_0,A_0=0))$, which equals $\E(Y^{0,0})$ under sequential conditional exchangeability, sequential positivity, and causal consistency. 

For $\theta(P)=\E(\E(\E(Y|W_0,A_0=0,W_1,A_1=0)|W_0,A_0=0))$, the EIF under a nonparametric statistical model is
\begin{align*}
    \phi'(O_i,P)&=\frac{\I(A_0=A_1=0)}{\Pr(A_0=0|W_0)\Pr(A_1=0|W_0,A_0=0,W_1)}\left[Y-\mu\right] \\
    &\quad\quad+ \frac{\I(A_0=0)}{\Pr(A_0=0|W_0)}\left[\mu-E(\mu|W_0,A_0=0)\right] \\
    &\quad\quad+ E(\mu|W_0,A_0=0)-\psi
\end{align*}
where $\mu=E(Y|W_0,A_0=0,W_1,A_1=0)$.

The estimating equation of the EIF, $0=\sum_{i=1}^n \phi'(O_i,P)$ can be expressed as a set of 3 estimating equations. Again, as in the main text, we use a $^*$ to distinguish the targeted versions of $\mu$ and $E(\mu|W_0,A_0=0)$. 
\begin{align}
    0 &= \sum_{i=1}^n\frac{\I(A_0=A_1=0)}{\widehat{\Pr}(A_0=0|W_0)\widehat{\Pr}(A_1=0|W_0,A_0=0,W_1)}\left[Y-\widehat{\mu}^*\right] \label{eq:app1}\\
    0 &= \sum_{i=1}^n \frac{\I(A_0=0)}{\widehat{\Pr}(A_0=0|W_0)}\left[\widehat{\mu}^*-\widehat{\E}^*(\widehat{\mu}^*|W_0,A_0=0)\right] \label{eq:app2}\\
    0 &= \sum_{i=1}^n \widehat{\E}^*(\widehat{\mu}^*|W_0,A_0=0)-\psi \label{eq:app3}
\end{align}

Equation (\ref{eq:app3}) can be rearranged into a closed-form estimator of our estimand, $\widehat\theta_{target} = \frac{1}{n}\sum_{i=1}^n \widehat{\E}^*(\widehat{\mu}^*|W_0,A_0=0)$. Thus we can estimate our estimand once we have $\widehat{\E}^*(\widehat{\mu}^*|W_0,A_0=0)$, which we can estimate by first solving (\ref{eq:app1}) and then solving (\ref{eq:app2}).

First consider (\ref{eq:app1}). We can define $\widehat{\mu}^*=\widehat{\mu} + \widehat \gamma_0$ and plug this into (\ref{eq:app1}) to get the score equation 
$$0 = \sum_{i=1}^n\frac{\I(A_0=A_1=0)}{\widehat{\Pr}(A_0=0|W_0)\widehat{\Pr}(A_1=0|W_0,A_0=0,W_1)}\left[Y-\left\{\widehat{\mu} + \widehat \gamma_0\right\}\right].$$ This score equation is solved by estimating, using maximum likelihood, an intercept-only linear model of $Y$ with $\widehat\mu$ as an offset with individual weights $\frac{\I(A_0=A_1=0)}{\widehat{\Pr}(A_0=0|W_0)\widehat{\Pr}(A_1=0|W_0,A_0=0,W_1)}$. Predictions from this fitted model are $\widehat{\mu}^*$.

Next consider (\ref{eq:app2}). We can define 
$\widehat{\E}^*(\widehat{\mu}^*|W_0,A_0=0)=\widehat{\E}(\widehat{\mu}^*|W_0,A_0=0) + \widehat \gamma_1$ 
and plug this into (\ref{eq:app2}) to get the score equation 
$$0=\sum_{i=1}^n \frac{\I(A_0=0)}{\widehat{\Pr}(A_0=0|W_0)}\left[\widehat{\mu}^*-\left\{\widehat{\E}(\widehat{\mu}^*|W_0,A_0=0) + \widehat \gamma_1\right\}\right].$$ 
This score equation is solved by estimating, using maximum likelihood, an intercept-only linear model of $\widehat{\mu}^*$ with $\widehat{\E}(\widehat{\mu}^*|W_0,A_0=0)$ as an offset with individual weights $\frac{\I(A_0=0)}{\widehat{\Pr}(A_0=0|W_0)}$. Predictions from this fitted model are $\widehat{\E}^*(\widehat{\mu}^*|W_0,A_0=0)$.

Thus we can use the following algorithm for estimation
\begin{enumerate}
    \item Propensity score models: Fit flexible models for the treatment at each time point, possibly using machine learning, to estimate $\widehat{\Pr}(A_0=0|W_0)$ and $\widehat{\Pr}(A_1=0|W_0,A_0=0,W_1)$
    \item Outcome model at time point 1: Fit a flexible model for $\E(Y|W_0,A_0=0,W_1,A_1=0)$, possibly using machine learning. Predictions from this model are $\widehat \mu$
    \item Targeting model at time point 1: Fit an intercept-only linear model of $Y$ with $\widehat\mu$ as an offset, weighted by $\frac{\I(A_0=A_1=0)}{\widehat{\Pr}(A_0=0|W_0)\widehat{\Pr}(A_1=0|W_0,A_0=0,W_1)}$, using maximum likelihood. Predictions from this model are $\widehat{\mu}^*$.
    \item Outcome model at time point 0: Fit a flexible model for $\E(\widehat{\mu}^*|W_0,A_0=0)$, possibly using machine learning. Predictions from this model are $\widehat{\E}(\widehat{\mu}^*|W_0,A_0=0)$
    \item Targeting model at time point 0: Fit an intercept-only linear model of $\widehat{\mu}^*$ with $\widehat{\E}(\widehat{\mu}^*|W_0,A_0=0)$ as an offset, weighted by $\frac{\I(A_0=0)}{\widehat{\Pr}(A_0=0|W_0)}$, using maximum likelihood. Predictions from this model are $\widehat{\E}^*(\widehat{\mu}^*|W_0,A_0=0)$.
    \item Estimate estimand as the sample mean of $\widehat{\E}^*(\widehat{\mu}^*|W_0,A_0=0)$
\end{enumerate}

As in the example in the main text, we could have alternatively considered targeting models where $\frac{\I(A_0=A_1=0)}{\widehat{\Pr}(A_0=0|W_0)\widehat{\Pr}(A_1=0|W_0,A_0=0,W_1)}$ and $\frac{\I(A_0=0)}{\widehat{\Pr}(A_0=0|W_0)}$ are included as covariates. Further, as discussed in the main text, although we have shown linear targeting models to simplify notation, logistic models are recommended for better finite sample properties.
